\documentclass[a4paper]{book}
\usepackage{epsf}
\usepackage{nano2cmr}
\usepackage{textcomp}

\begin{document}

\pnum{}
\ttitle{STM investigation of structural properties of Si layers deposited on Si(001) vicinal surfaces}
\tauthor{{\em L.~V.~Arapkina}, V.~A.~Chapnin, K.~V.~Chizh, L.~A.~Krylova, V.~A.~Yuryev}

\ptitle{STM investigation of structural properties of Si layers deposited on Si(001) vicinal surfaces}
\pauthor{{\em L.~V.~Arapkina}$^{1}$, V.~A.~Chapnin$^{1}$, K.~V.~Chizh$^{1}$, L.~A.~Krylova$^{1}$,  V.~A.~Yuryev$^{1,2}$}

\affil{$^{1}$~Prokhorov General Physics Institute of the Russian Academy of Sciences, 38 Vavilov Street, Moscow, 119991, Russia
\\ $^{2}$~Technopark of GPI RAS, 38 Vavilov Street, Moscow, 119991, Russia}

\begin{abstract}
{This report covers investigation of the structural properties of surfaces of Si epitaxial layers deposited on Si(001) vicinal substrates with different miscuts. We have shown processes of generation and growth of surface defects to depend on tilt direction of a Si(001) wafer and epilayer growth mode. We suppose these effects to be connected with interaction of monoatomic steps.}
\end{abstract}

\begindc 

\index{Yuryev V. A.}
\index{Arapkina L. V.}   
\index{Krylova L. A.}
\index{Chapnin V. A.}
\index{Chizh K. V.}

\section*{Introduction}

A structure of a Si(001) epitaxial layer surface, especially its defects, could affect the formation of nanostructures. Perfect, defectless Si epilayers grown on Si(001) vicinal substrates are of special importance for such industrially significant problem as controllable formation of Ge/Si(001) nanostructures for optoelectronic device applications. This report covers experimental investigation of structural properties of surfaces of Si epitaxial layers deposited on Si(001) vicinal surfaces of substrates with different miscuts.

\section {Experimental}

Experiments were carried out in UHV using GPI-300 STM coupled with Riber EVA 32 MBE chamber [1].
Epitaxial layers were deposited by MBE on substrates cut from Si(001) vicinal wafers tilted $\sim$\,0.2{\textdegree} towards the [110] or [100] direction. Initial surfaces were treated by the RCA etchant. Before Si deposition, we cleaned the surfaces by the standard methods of preliminary annealing at 600{\textcelsius} and decomposition of the SiO$_2$ film under a weak flux of the Si atoms at 800{\textcelsius} [2].
The substrate temperature during Si deposition was chosen in the range from 360 to 700{\textcelsius}.



\section{STM data}

We have explored structural properties of Si epitaxial films deposited on Si(001) vicinal substrates depending on the growth temperature and the rate of Si deposition. Two modes of Si epitaxial growth have been observed. We have found that the step-flow growth goes on at the temperatures above 600{\textcelsius} whereas the island growth takes place  at the temperatures be\-low 600{\textcelsius}. Samples grown at the step-flow growth mode have  smooth surfaces composed of terraces bounded by mono\-atomic steps. STM data  shown in Figs.~1
and 2
are related to the Si epilayers deposited on the wafers tilted towards the [110] and [100] direction respectively. 

\subsection{Surfaces tilted towards [110]}

At first, we consider STM data for Si/Si(001) surfaces  tilted towards the [110] direction. The surface is composed by  S$_{\rm A}$ and S$_{\rm B}$ mono\-atomic steps [3];
S$_{\rm B}$ steps are wider than S$_{\rm A}$ ones.  We have observed formation of such defects as faceted pits on these surfaces. In Fig.~1,
an initial stage of the defect formation is shown. There observed the local stoppage of growth of an S$_{\rm B}$ step and appearance of two S$_{\rm A}$ steps instead. In other words, we have observed a gap of the S$_{\rm B}$ step. The S$_{\rm A}$ steps repulse each other and the defect cannot be overgrown quickly. The bottom of the deep pit has a rectangular shape and a long side of it formed by the S$_{\rm A}$ step. 

\begin{figure}[b]
\leavevmode
\centering{\epsfbox{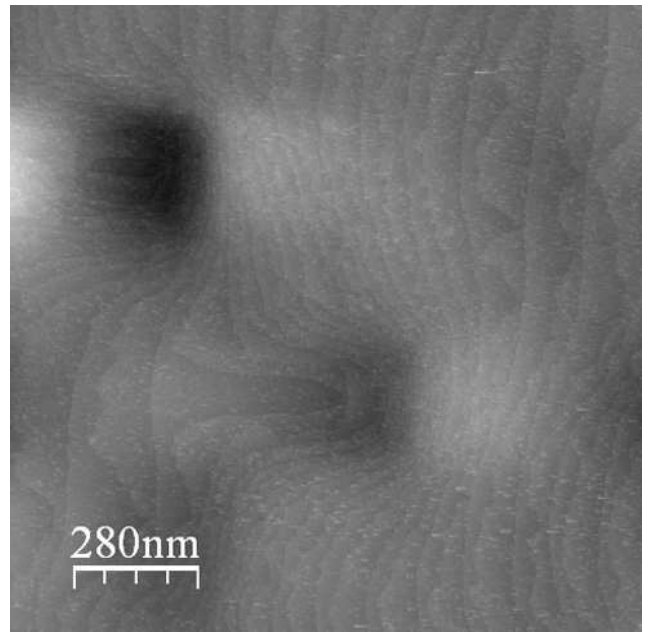}}
\caption{
STM image of the surface of the 50-nm Si epilayer deposited at 650{\textcelsius} on the Si(001) vicinal wafer tilted $\sim$\,0.2{\textdegree} towards [110], the deposition rate was $\sim$\,0.3\,\AA/c.}
\end{figure}

\begin{figure}[ht]
\leavevmode
\centering{\epsfbox{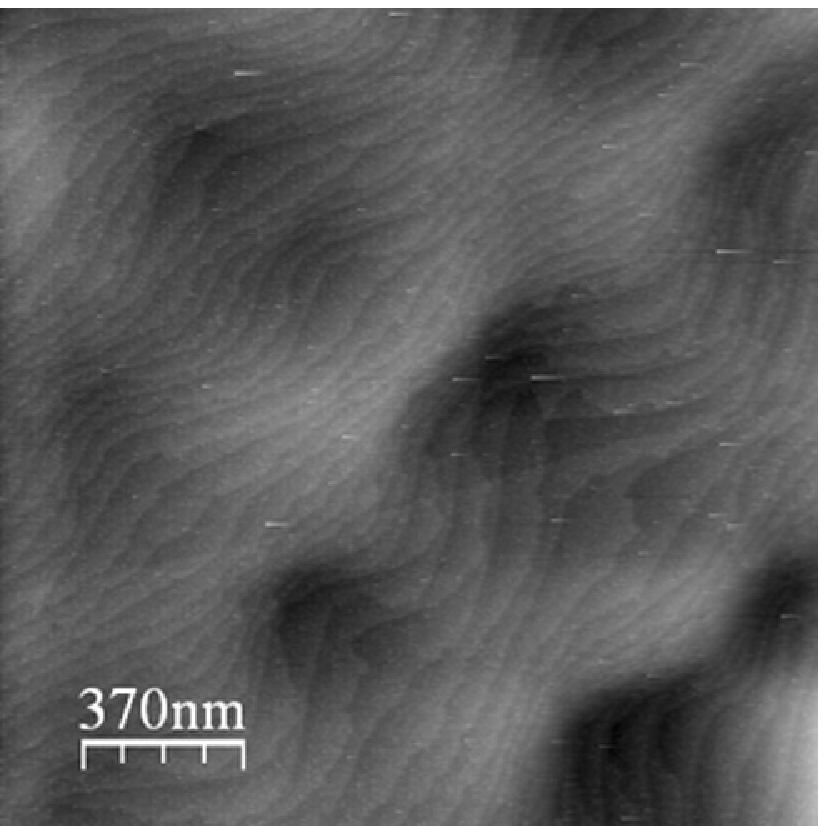}}
\caption{
STM image of the surface of the 50-nm Si epilayer deposited at 650{\textcelsius} on the Si(001) vicinal wafer tilted $\sim$\,0.2{\textdegree} towards [100], the deposition rate was $\sim$\,0.3\,\AA/c.}
\end{figure}

\begin{figure}[ht]
\leavevmode
\centering{\epsfbox{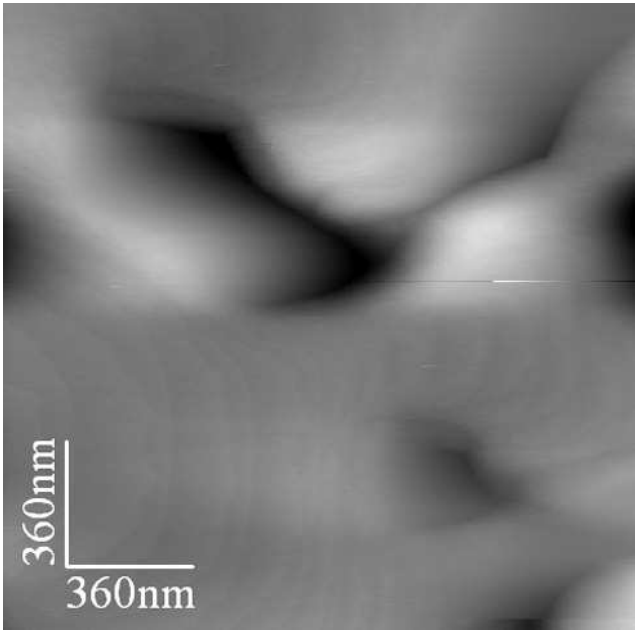}}
\caption{
STM image of the surface of the 50-nm Si epilayer deposited at 650{\textcelsius} on the Si(001) vicinal wafer tilted $\sim$\,0.2{\textdegree} towards [100], the deposition rate was $\sim$\,0.1\,\AA/c.}
\end{figure}

\subsection{Surfaces tilted towards [100]}

STM data for the Si/Si(001) surface  tilted towards the  [100] direction are presented in Fig.~2.
The surface is composed by bent mono\-atomic steps. In this case every mono\-atomic step consists of  short parts of S$_{\rm A}$ and S$_{\rm B}$ steps and runs along the [100] direction. There are local disarrangements of the structure. We suppose this kind of defects to be connected with a process of transition from the mixed S$_{\rm A}$\,+\,S$_{\rm B}$ monoatomic step to two single S$_{\rm A}$ and S$_{\rm B}$ steps instead of formation of the D$_{\rm B}$ step. 

We have  investigated the structural properties of the Si film surface depending on the rate of Si deposition. We have found that reduction of the  Si deposition rate from ~0.3\,\AA/c (Fig.~2)
to 0.1\,\AA/c (Fig.~3)
results in appearance of the structure formed by monoatomic steps  running along the [110] direction and formation of shapeless pits on the surface instead of the structure formed by the bent monoatomic steps which run along the [100] direction. 

\subsection{Effect of temperature reduction}

In Figs.~4
and 5,
we present STM data for surfaces of the epilayers deposited at the temperatures of 550 and 470{\textcelsius}.

 Surfaces of the samples grown at 550{\textcelsius} consist of  S$_{\rm A}$ and S$_{\rm B}$ mono\-atomic steps (Fig.~4).
The mixed S$_{\rm A}$\,+\,S$_{\rm B}$ monoatomic steps which are typical for the Si(001) surfaces tilted towards [100] and obtained at higher temperatures are not observed. At this temperature, the transitional  mode of the epilayer growth, intermediate between the island growth and the step-flow one, is observed for both tilt directions. 

Further reduction of the temperature down to 470{\textcelsius} results in the island growth mode (Fig.~5).
The Si/Si(001) surface is composed by small islands. In both causes  such defects as pits are present on the surface. The structural properties of the surfaces of  Si epilayers grown at the island growth mode do not depend on the direction of the surface tilt.

\begin{figure}[t]
\leavevmode
\centering{\epsfbox{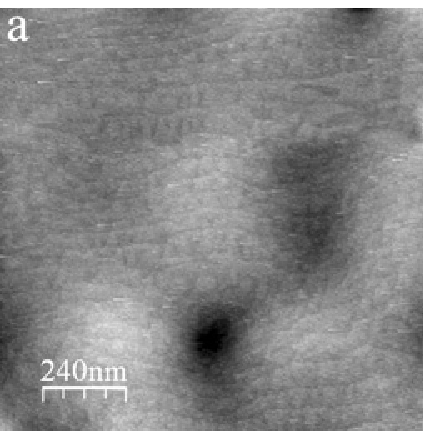}}
\centering{\epsfbox{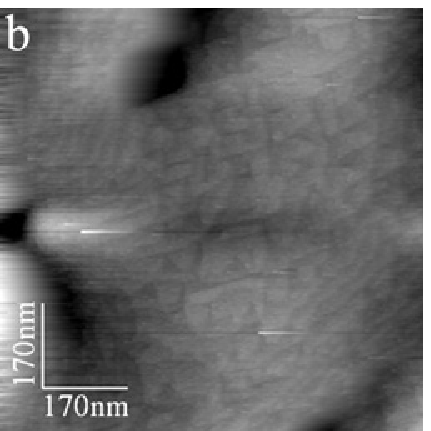}}
\caption{
STM image of the surface of the 50 nm thick Si epilayer deposited at 550{\textcelsius} on the Si(001) vicinal wafer (deposition rate is $\sim$\,0.3\,\AA/c);  tilt is $\sim$\,0.2{\textdegree} towards (a)  [110] and (b) [100].}
\end{figure}

\begin{figure}[t]
\leavevmode
\centering{\epsfbox{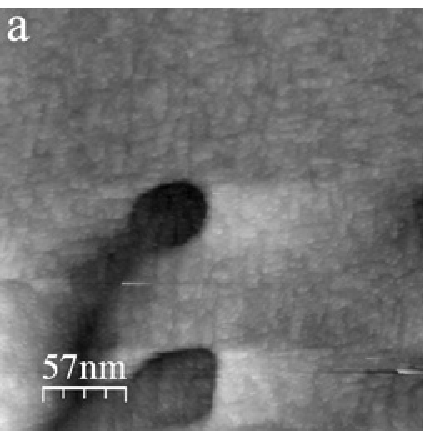}}
\centering{\epsfbox{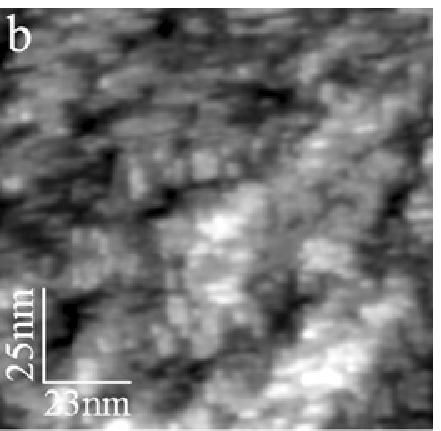}}
\caption{
STM image of the surface of the 50 nm thick Si epilayer deposited at 470{\textcelsius} on the Si(001) vicinal wafer (deposition rate is $\sim$\,0.3\,\AA/c);  tilt is $\sim$\,0.2{\textdegree} towards (a)  [110] and (b) [100].}
\end{figure}

\section{Conclusion}

 Summarizing the above we can conclude that processes of generation and growth of surface defects arising during epitaxial growth of Si films on Si(001) vicinal substrates depend on tilt direction of a Si(001) wafer and the epilayer growth conditions. We suppose the  observed effects to be a consequence of mutual interactions of monoatomic steps.

\acks  This research has been supported by the Ministry of Education and Science of Russian Federation through the contracts No.~14.740.11.0069 and 16.513.11.3046. Facilities of Center of Collective Use of Scientific Equipment of GPI RAS were utilized in this research. We appreciate the  financial and technological support.



\begin{thebibliography}{8}
\itemsep-2pt






\bibitem{ara1-NRL2011-VCIAN} 
V.~A.~Yuryev \etal, {\em Nanoscale Res. Lett.} {\bf 6}, 522 (2011).

\bibitem{ara1-jetpl-phase-trans} 
L.~V.~Arapkina \etal, {\em JETP Lett.} {\bf 92}, 310 (2010).

\bibitem{ara1-Chadi-steps} 
D.~J.~Chadi, {\em Phys. Rev. Lett.} {\bf 59}, 1691 (1987).



\end{thebibliography}
\end{document}